\documentstyle[11pt,psfig,aaspp4]{article}
\begin{document}

\title {CROSSCORRELATION SPECTROPOLARIMETRY IN SINGLE-DISH RADIO ASTRONOMY}

\author{ Carl Heiles}
\affil {Astronomy Department, University of California,
    Berkeley, CA 94720-3411; cheiles@astron.berkeley.edu}

\begin{abstract}

	Modern digital correlators permit the simultaneous measurement
of all four Stokes parameters using auto and crosscorrelation.  We
briefly outline the fundamental requirements and some practical details
of performing such measurements and refer to two additional papers
that provide overview and cover calibration issues in detail. 

\end{abstract}


\section {INTRODUCTION TO CROSSCORRELATION SPECTROPOLARIMETRY}
\label{intro}

	Modern digital correlators permit complete, simultaneous
spectropolarimetric measurements by providing the frequency spectra of
all four Stokes parameters using auto and crosscorrelation.  This paper
is a very brief introduction to two additional papers (Heiles et al
2001a, 2001b) that provide detailed explanations of the calibration
issues of such measurements.  One of these deals with the measurement
and calibration of instrumental effects on Stokes parameters using the
$4 \times 4$ Mueller matrix.  The other deals with the more general
issue of parameterizing the telescope beam and first sidelobe, for not
only the usual Stokes $I$ but also the other three Stokes parameters;
beam polarization effects can lead to serious instrumental effects when
measuring extended emission. 

	Our work is based on experience with systems at three
telescopes: the Spectral Processor at the NRAO 140 foot
telescope\footnote{The National Radio Astronomy Observatory is a
facility of the National Science Foundation operated under cooperative
agreement by Associated Universities, Inc.}, the interim correlator at
the Arecibo telescope\footnote{The Arecibo Observatory is part of the
National Astronomy and Ionosphere Center, which is operated by Cornell
University under a cooperative agreement with the National Science
Foundation}, and the large digital correlator at the NRAO 12-m
telescope.  We learned about instrumental effects that can affect the
astronomical measurements unless they are properly calibrated and
equipment details that need to be accounted for in the calibration. 

	In the present short paper we briefly outline the the technique
and cover a few practical details of performing such measurements.  Our
purpose here is not to present a complete overview or to reiterate the
many important details discussed at length in the other papers.  Rather,
it is to present the briefest of brief introductions and, in addition,
to cover several practical details for the specialist.  The basic
reference for our work is the excellent book on astronomical
polarization by Tinbergen (1996).  A more mathematical and fundamental
reference is Hamaker, Bregman, and Sault (1996), which the
theoretically-inclined reader will find of interest. 

	The four Stokes parameters define the complete state of
polarization of electromagnetic radiation.  They are most simply thought
of as linear combinations of orthogonal polarizations.  Stokes $I$, the
total intensity, is the sum of any two orthogonal polarizations.  Stokes
$Q$ and $U$ define the linear polarization: $Q = i_0 - i_{90}$ and $U =
i_{45} - i_{-45}$, where $i$ is the power measured in a particular
polarization and the numerical subscripts define the position angle of
the received linear polarization.  Stokes $V$ defines the circular
polarization: $V = i_{LCP} - i_{RCP}$.  Here we follow the IEEE
definition for circular polarization; $LCP$ means left-hand-circular
polarization.  $LCP$ is generated by transmitting with a left-handed
helix, so from the transmitter the $E$ vector appears to rotate
anticlockwise.  From the receiver, $LCP$ appears to rotate clockwise. 
Generally in this paper we follow Tinbergen (1996), whose definition of
$LCP$ is the same as ours; however, his definition of $V$ (see his page
61) is opposite ours and, also, the conventional one used by radio
astronomers. 

	More generally, the Stokes parameters can be written as
time-average products of the electric fields. In a radio astronomy feed
these electric fields are converted to voltages, which are equivalent. 
Let us focus on the specific case of two orthogonal linear polarizations
that provide voltages $(E_X, E_Y)$. Here the directions  $(X,Y)$ specify
the coordinate system with respect to which  $PA_{src}$, the
polarization position angle of an astronomical source, is measured; by
astronomical convention, $PA_{src}$ is measured with respect to the
North Celestial Pole, increasing towards the East, so $X$ points North
and $Y$ points East. 

	For a monochromatic signal, the voltages $(E_X, E_Y)$ vary
sinusoidally with time and can be written in complex notation, for
example $E_X = E_{X0} e^{i(2\pi f t + \psi_x)}$, where $f$ is the
observing frequency and $\psi$ is a phase angle. An arbitrary signal is
a sum of such monochromatic signals and is also complex. We then have

\begin{mathletters}
\label{Sdefinition1}
\begin{equation}
I = E_X \overline{ E_X} + E_Y \overline{ E_Y}
\end{equation}
\begin{equation}
Q = E_X \overline{ E_X} - E_Y \overline{ E_Y}
\end{equation}
\begin{equation}
U = E_X \overline{ E_Y} + \overline{ E_X} E_Y
\end{equation}
\begin{equation}
iV = E_X \overline{ E_Y} - \overline{ E_X} E_Y \; .
\end{equation}
\end{mathletters}

\noindent where the bars indicate complex conjugate and it is understood
that the products are averaged over time.  These time average products
can be obtained by suitable equipment that multiplies the voltages.  In
practice the products are performed digitally, with the real and
imaginary portions being obtained by mixing the incoming signal with a
local oscillator having $0^\circ$ and $90^\circ$ phase shifts. 

	To obtain spectral information, there are two almost equivalent
techniques. More usual in single-dish work is the XF technique: one
calculates time-average correlation functions digitally (the X part) and
takes the Fourier transform (the F part). The alternative is the FX
technique, in which one calculates Fourier transforms (F), multiplies by
the complex conjugate (X), and averages the resulting spectra. According
to the convolution theorem, these produce identical results. However, it
is not generally appreciated that the convolution theorem applies only
if the sampling time is infinite. For real data with a finite sampling
time, the off-frequency spurious response, sometimes called ``spectral
leakage'', differs for the two techniques.  For example,  consider the
example of a monochromatic signal of frequency $f_0$. In the XF case,
the spectral leakage has the form $ \sin x \over x$, where $x \propto (f
- f_0)$, and the power spectrum can be negative. In contrast, with FX
processing the power spectrum is the product of a Fourier transform with
its own complex conjugate, and is therefore everywhere positive. For the
XF technique the spectral leakage can be reduced by appropriate
convolution of the power spectrum, which is equivalent to weighting the
time window, e.g.\ with Hanning weighting. However, with the FX
technique convolution doesn't work and the window weighting must be done
on the Fourier transformed voltages before the multiplication.

	The Green Bank Spectral Processor uses the FX technique; its
output is time-average products of Fourier transforms. In
equations~\ref{Sdefinition1}, one should replace the voltage products 
by these products of Fourier transforms. For example, $E_X
\overline{E_X}$ is replaced by $FT(E_X) \overline{ FT(E_X)}$ and $E_X
\overline{E_Y}$ is replaced by $FT(E_X) \overline{ FT(E_Y)}$, where $FT$
means Fourier transform. The $FT$'s are complex, so while the self
products [like $FT(E_X) \overline{ FT(E_X)}$] are always real, the cross
products [like $FT(E_X) \overline{ FT(E_Y)}$] are complex.

	Standard digital correlators, like the Arecibo and NRAO
correlators, use the XF technique; the outputs are time-average
correlation functions. In equations~\ref{Sdefinition1}, one should
replace the voltage products by Fourier transforms of these correlation
functions. For a single signal like $E_X$, the correlation function is
the autocorrelation function. Autocorrelation functions are symmetric
about zero time lag and therefore the Fourier transforms are real; these
produce Stokes $I$ and $Q$ above. In contrast, for the two signals of
orthogonal polarizations $E_X$ and $E_Y$, the correlation function is
the crosscorrelation function.  The cross correlation function has
arbitrary symmetry so its Fourier transform is complex; the real part is
$U$ and the imaginary one $V$.  These matters are discussed at length in
books on interferometry, e.g.~Thompson, Moran, \& Swenson (1994).  

	Equations \ref{Sdefinition1} assume that the voltages are
properly calibrated for polarization.  ``Calibration for polarization''
means, specifically, that the relative gains and phases of the two
polarization channels are measured and the differences corrected for. 
Heiles et al (2001a) explains these calibration matters in detail.  An
additional aspect of calibration is the establishment of an accurate
scale of antenna temperature or flux density; this aspect is beyond the
scope of the current discussion, which focuses on polarization
calibration. 

	Two quantities, the relative gain and phase of the two
polarization channels, must be calibrated.  In principle, a good way to
do this is with an astronomical source having known linear polarization.
 However, in practice this is not sufficient because the phase is
ambiguous by $180^\circ$ (see \S\ref{ambig}) and, moreover, the phase
delay can change with telescope position, e.g.~when the i.f.~cables are
mechanically strained at the telescope bearings.  Therefore, a secondary
calibration standard that is attached to the telescope, one that can be
used quickly without moving the telescope, is required.  This standard
is best realized by injecting a correlated noise source into the two
feed outputs or, alternatively, by radiating a noise source into the
feed with a polarization that provides equal amplitudes in the feed
outputs.  Under some circumstances it is desirable to inject noise
separately into the two inputs to simulate uncorrelated noise; these
matters are discussed in detail by Heiles and Fisher (1999). 

	In the following sections we describe several practical details
associated with the technique of single-dish crosscorrelation
spectropolarimetry. These details go beyond the calibration techniques
conventionally required for single-dish radio astronomy, such as are
discussed by Rohlfs and Wilson (1999), for example. Some of these
details are familiar to practitioners of interferometric
crosscorrelation spectropolarimetry; however, we suspect that persons
who concentrate on single-dish techniques may find them unfamiliar (as
we did). 

\section{ THE RELATIVE PHASE {\boldmath $\psi$}}

	We define $\psi$ to be the relative phase between the two
polarization channels that enter the correlator.  It consists of two
parts, the source contribution $\psi_{src}$ and the instrumental one
$\psi_{sys}$. Thus,

\begin{equation}
\psi = \psi_{src} + \psi_{sys} \ .
\end{equation}

\noindent $\psi_{src} = 0^\circ$ or $180^\circ$ for a linearly polarized
source and $\pm 90^\circ$ for a circularly polarized one. $\psi_{sys}$
is produced by several effects as discussed below.

\subsection{ TWO COMMON MISCONCEPTIONS CONCERNING {\boldmath $\psi_{src}$}}
\label{ambig}

	Suppose for the moment that $\psi_{sys}=0^\circ$ and, further,
that the feed has orthogonal linear polarizations oriented at position
angles $(0^\circ, 90^\circ)$.  A common misconception is that the
observed $\psi$ from a linearly-polarized source depends on the position
angle so that, for example, it changes smoothly with the parallactic
angle as the source is tracked with an alt-az telescope.  Another is
that the observed $\psi$ is independent of position angle. Neither is
the case. 

	The first misconception arises because one erroneously imagines
that the relative {\it amplitude} of $E_X$ and $E_Y$ determine $\psi$;
this amplitude ratio changes with position angle.  However, $\psi$ is
the {\it phase} difference between $E_X$ and $E_Y$.  If $U$ is positive,
this phase is $\psi=0^\circ$; if $U$ is negative, $\psi=180^\circ$.  $U$
changes sign when the position angle crosses the boundaries $0^\circ$
and $90^\circ$ (recall that position angle is modulo $180^\circ$, so
each crossing occurs twice in a full rotation). 

	It is important to remember that the relative phase of a
linearly-polarized source is bimodal, with the two values $0^\circ$ and 
$180^\circ$.  It means that without keeping track of the position angle,
one cannot rely on an astronomical source to calibrate the instrumental
relative phase because of the $180^\circ$ ambiguity. 

\subsection{ THE EFFECT OF AMPLIFIERS AND SIDEBAND ON {\boldmath $\psi_{sys}$}}

	One needs to adjust the signal level entering the correlator to
its optimum level. This is accomplished with a combination of amplifiers
and attenuators, and often this combination must be changed in response
to the source intensity. Many types of amplifier introduce $180^\circ$
phase shifts. If one changes the number of amplifiers in the chain, then
the phase can jump by multiplies of $180^\circ$; this can be
disconcerting. Of course, when switching in attenuators and amplifiers,
the path lengths usually change, and this also produces phase changes. 

	In a similar vein, we found that most of the contribution to
$\psi_{sys}$ comes from the i.f.~cables. However, we always refer $\psi$
to the r.f.~frequency. For that part of $\psi_{sys}$ contributed at any
i.f.~stage, the sign of its $\psi$ depends on whether that stage
processes the upper or lower sideband.

\subsection{ A PRACTICAL DETAIL IN PHASE CALIBRATION: A ROBUST LEAST
SQUARE FIT OF THE PHASE VERSUS FREQUENCY} \label{relphase4}

	The relative phase delay caused by the system, $\psi_{sys}$, can
have several causes, but in our experience the most important is
differences in cable lengths between the two channels.  These
differences can occur at r.f.~(before the first mixer) and i.f.~(after
the first mixer).  For most telescopes the latter dominates, probably
because there is usually a long cable run from the feed to the control
room. Not always, however: at the NRAO 12-m telescope, the feeds for the
two polarization channels are mounted separately so that the incoming
signals have much different path lengths; moreover, the mountings are
mechanically soft so that this path length changes with telescope
position. 

	The difference in cable length produces a linear phase
difference with frequency given by

\begin{equation}
{d \psi_{sys} \over df} = {2 \pi \Delta L \over c} \ ,
\end{equation}

\noindent where $\Delta L$ is the difference in electrical length and
$\psi_{sys}$ the instrumental phase difference between the two channels.
 For example, at Arecibo we find ${d \psi_{sys} \over df} \sim 0.1$ rad
MHz$^{-1}$, which corresponds to a length difference of $\sim 5$ m, most
of which probably occurs along the pair of $\sim 500$ m optical fibers
that carry the two channels from the feed to the control room. 

	The quantity $\Delta L \over c$ is just the time difference
between the two signal paths. This suggests that one could obtain ${d
\psi_{sys} \over df}$ by injecting correlated noise at the front end and
locating the time offset from zero delay of the peak of the
crosscorrelation function. However, this doesn't work in practice, for
two reasons. First, the time delay is usually a tiny fraction of the
sampling time. For example, at Arecibo the time delay arising from the 5
meter cable length difference is of order 0.01 microsec, while often our
observing bandwidth is about 1 MHz so the sampling time is of order 1
microsec.  Thus the time delay is about $1\%$ of the sampling interval.
Locating a peak to $1\%$ of the sampling interval requires a difficult,
ill-defined nonlinear least squares fit. Second, the two versions of
signal come through entirely different signal paths, namely the two
polarization channels, and typical single dish electronics are usually
not engineered to optimize crosscorrelation so that the filter shapes in
the two paths are not very well matched. This can make the
crosscorrelation function rather complicated and even obliterate the
expected well-defined single peak. 

	In contrast, the phase slope is always very well defined and,
moreover, it is precisely the quantity required for calibrating the
polarization characteristics.  One measures this by injecting a
correlated noise source (``cal'') at the front end and fitting the
linear frequency dependence of the phase difference.

	Such fits are best done using the least-squares technique. One
might be tempted to make a linear polynomial fit of $\psi$ to frequency.
However, this is much more difficult than it seems because $\psi$
suffers sudden wraparound jumps of $2 \pi$ when it crosses the
boundaries $-\pi$ or $\pi$. Moreover, $\psi$ has noise, and this makes
the locations of these jumps difficult to determine. One can surmount
this difficulty by various {\it ad hoc} subterfuges, but there is a
better way.

	First, realize that we define the phase $\psi$ by calculating

\begin{equation}
\psi = \tan^{-1} \left[ {\rm Im}(E_XE_Y) \over {\rm Re} ( E_XE_Y)
   \right] \; ,
\end{equation}

\noindent which means that $\sin \psi = {\rm Im}(E_XE_Y)$
and $\cos \psi = {\rm Re} (E_XE_Y)$. We write

\begin{equation}
\psi = A + B f
\end{equation}

\noindent where $f$ is the r.f.~frequency and $(A,B)$ are the constants
we need to determine.  Write 

\begin{mathletters}
\label{newfit}
\begin{equation}
\sin (A + Bf) = S\!A \cos(Bf) + C\!A \sin(Bf) = {\rm Im} (E_XE_Y) \ ,
\end{equation}
\begin{equation}
\cos (A + Bf) = C\!A \cos(Bf) - S\!A \sin(Bf) = {\rm Re} (E_XE_Y)
\end{equation}
\end{mathletters}

\noindent where $(C\!A,S\!A) = (\cos A, \sin A)$.  Now determine $(C\!A,
S\!A, B)$ using least squares. A very robust procedure is to use a
reasonably accurate guess for $B$ and simultaneously fit for $(C\!A,
S\!A)$ in equation~\ref{newfit}; this is a straightforward linear least
squares fit. Then use the results as input estimates in a full
three-parameter nonlinear least squares fit.

\section{  A PRACTICAL DETAIL IN FOURIER TRANSFORMING CROSSCORRELATION
FUNCTIONS}

	One generally uses the FFT algorithm when Fourier transforming
correlation functions, with the number of channels equal to an integral
power of 2.  Some care is required in taking such Fourier transforms of
correlation functions. 

	Digital correlators for single dishes are usually engineered to
provide positive time lags.  However, crosscorrelation requires both
positive and negative lags.  One obtains the full crosscorrelation
function of $(E_X, E_Y)$ by measuring {\it two} correlation functions
with positive lags, one delaying $E_Y$ and the other $E_X$.  These must
be combined before taking the Fourier transform. 

	The process of combining the two correlation functions to
produce a crosscorrelation function (and that of symmetrizing an
autocorrelation function) runs into a particular practical detail.  The
two halves of a crosscorrelation function each have $N$ channels, with
time lags running from $0 \rightarrow (N-1)\tau$ and $0 \rightarrow
-(N-1)\tau$, because correlators are always engineered so that both
halves measure the zero lag product.  (Here $\tau$ is the sampling time
interval). Thus the total number of independent channels is not $2N$,
but rather $2N-1$.  However, the FFT algorithm requires $2N$ input
numbers.  

	We have a ``missing channel''.  This missing channel has a time
lag equal $N\tau$, and also equal to $-N\tau$: from the fundamental
assumptions inherent in digital Fourier transforms, the correlation
values for the $\pm N\tau$ must be equal.  One must set the unknown
correlation value for this channel to a reasonable number.  The proper
choice for this number is important only insofar as it should produce no
discernible impact on the derived power spectrum. 

	Generally, correlation functions for random noise tend towards
zero at large lags.  The unknown channel has a large lag, so one might
be tempted to set its value to zero.  This is the wrong choice! There
are two reasons: one, the signal may have some low-frequency components
that make its correlation at large lags nonzero; two, the digital
correlator may have a d.c.~offset.  In either case, setting the missing
channel equal to zero produces a spike in the correlation function,
which is reflected in the power spectrum as a channel-to-channel
oscillation.  The proper choice for the missing channel is the average
of the two values for $\pm (N-1)\tau$, because this produces no
extraneous effects in the derived power spectrum. 

	Often one weights the time window to reduce spectral leakage, as
mentioned above. Most weighting schemes, for example Hanning weighting,
assign zero weight to the missing channel. With such weightings, the
missing channel doesn't matter. Weighting is often desirable,
particularly when there is interference, to reduce the spectral
ringing. 

\section{SUMMARY}

	We have provided a brief introduction to the technique of
digital spectropolarimetry in radio astronomy, together with 
fundamental references. We also include some comments and details based
on experience that should help a specialist to get started with this
technique.

\acknowledgements

        It is a pleasure to acknowledge helpful comments by the referee,
which led to significant clarifications. This work was supported in part
by NSF grant 95-30590 to CH.

\end{document}